# Experimental Investigation of Aortic Pressure Variations Following a Simulated Thoracic Impact

Ghassan Maraouch*, Curtis H. Horton, Joseph Fanaberia, Eduardo Malorni, Gian-Carlo Mignacca, Mark Cohen, Lyes Kadem

Department of Mechanical Industrial and Aerospace Engineering, Concordia University, Montreal, Québec, Canada

**(*): Corresponding Author:**

*Address:* Department of Mechanical, Industrial and Aerospace Engineering, Concordia University, 1455 de Maisonneuve Blvd. W., Montreal, Quebec, Canada, H3G 1M8
*Tel.:* (514) 848-2424 ext. 3143
*Fax:* (514) 848-3175
*Email:* lcfd@encs.concordia.ca

# ABSTRACT

Blunt traumatic aortic rupture is a heart injury that can occur in falls, automobile accidents, and sporting injuries involving impact to the thorax. Despite its severity and high morbidity rate, the research still does not provide a consistent description of the mechanism of rupture. In this study, a crash testing dummy with an *in vitro* pumping heart, 3D printed ribcage, and ballistic gel damping layer was developed to reproduce a realistic response to thoracic impact. Testing was performed using a standardized pendulum used for calibration of crash test dummies, with the location of impact being the middle of the sternum. Different impact severities were tested by adjusting the kinetic energy at impact with the initial height of the pendulum. Measurements of the dummy include instantaneous aortic pressure waveforms prior, during and following the impact. The results of this experiment show that aortic pressure experiences significant changes in magnitude during simulated impact. This work could help contribute towards a better understanding of the mechanisms leading to blunt traumatic aortic rupture and the development of preventative measures.

# INTRODUCTION

Damage to the aorta can result in drastic reduction in quality of life and can also result in death in many cases. One type of injury in particular that occurs in severe automobile accidents, known as blunt traumatic aortic injury (BTAI), still has a lot of uncertainties regarding the mechanisms involved [1] [2]. In terms of severity, BTAI is the second leading cause of death in regarding blunt injuries, preceded by head injuries [1] [3]. Its severity is serious enough that it usually leads to death in around 80% of cases following the accident, prior to receiving hospital care [1] [3] [4]. Smith et al. [3] determined that 77% of death occurs prior to hospital care, with a survival rate of 71% given an opportunity for treatment. Overall, their study has shown that BTAI had a mortality rate of 84% [3]. Regarding causes, Fabian et al. [5] have reported that 81% of BTAI were a result of motor vehicle accidents. Side impacts were responsible for 24% of BTAI and only 4% were due to a rear impact; the remaining 72% was a result of head-on impact [5]. Their study has also shown that males were more likely to suffer from BTAI, with a reported number of 199 males out of 274 (72.6%) patients [5]. Smith et al. [3] also had a dominant male sample in their study, with 49 out of 61 (80.3%) of patients being male. Greendyke [6] found that 16% of victims from automobile accidents had rupture of the aorta. BTAI is not restricted to vehicular accidents. Other types of injuries that can cause BTAI are sport injuries to the chest or free fall.

Sections of the aorta are typically referred to as the aortic root, the ascending aorta, the aortic arch and the descending aorta. Tears due to BTAI are found in most cases at the isthmus region, located on the aortic arch, distal to the left subclavian artery [3] [7]. Damage has also been found on the ascending aorta, descending aorta, aortic arch, abdominal aorta and in some cases, at multiple sites [3] [7]. Tears have been identified to typically be smooth and transverse to the aortic wall, but other types of ruptures have also been identified [7].

Interestingly, this type of injury occurs in events where there is not necessarily a blunt injury to the chest. Rather, in events such as car accidents, the inertia from rapid deceleration has been suggested as the cause for the rupture [1] [8] [9]. Mechanisms that have been proposed as a cause of the rupture are a combination of stretching, increase in hemodynamic pressure, the osseous pinch or the water-hammer effect [1] [8] [9] [10]. The lack of consensus for a mechanism is mostly due to injury not being reproduced experimentally [10] or that a certain mechanism is not valid for a specific location where tears have been observed [9].

The main objective of this original study is to contribute to a better fundamental understanding of the mechanisms leading to BTAI. For this purpose, a unique experimental facility dedicated to *in vitro* evaluation of BTAI is designed and the temporal evolution of aortic pressure in the ascending aorta is investigated following different simulated severities of frontal impacts.

## METHODS

Investigation of pressure changes during impact was done by designing an original *in vitro* crash test dummy (Fig. 1). Actuation of the system was done via a linear motor that mechanically pumped a silicone left ventricle with controllable motion waveform, frequency and stroke, all of which were controlled through a custom-made LabVIEW code. This allows the system to simulate multiple heart rates and change waveforms for calibration of the desired aortic pressure waveform. For consistency, it was decided that the heart rate experimented with would be set to 70 beats per minute.

A reservoir is kept above the dummy for filling of the system and pressurization. Valves were added right after the reservoir to the circuit to close the system. Closing the system to atmospheric allowed the system to be air pressurized to attain the desired working pressure for calibration of the dummy. This made the system compact, generating less losses throughout the system and more versatile for assembly at outside facilities. The working fluid was a mixture of water-glycerol with a volume ratio of 60:40. At room temperature, this mixture has a density of 1090 kg/m$^3$ and a viscosity of approximately 4.2 cP. A combination of vinyl tubes and silicone rubber tubes were incorporated in the system. Vinyl tubes were mostly used where rigidity was needed as silicone rubber tubes would collapse or kink.

Transparent silicone (Polycraft Silastic T4) was used to make the models of the left ventricle and aorta. Manufacturing of the heart components was done by 3D-printing positive molds from CAD models designed to replicate the anatomical structure of the left ventricle and aorta. Silicone was then made in batches for each layer and brushed on the mold. Preparation of the silicone layers consists of mixing the silicone with a curing agent at a mass ratio of 10:1 and then placed in a vacuum for the removal of air bubbles. To control the layer thickness and cure faster, the molds were placed in a custom-made heated chamber with a rotating machine that evens out the layers. Each layer took approximately 1 hour to cure, and 4 to 5 layers were added, depending on the thickness desired.

A thoracic cage made of 3D printed PLA with layer height of 0.1mm and 100% infill was used to encase the ventricle and aorta. Printing of the ribs, sternum and spine were done separately and assembled via nuts and bolts. Fastening of the ribs gave more leeway for assembly of the system and the option to change individual ribs if they have any defects. A layer of ballistic gel was placed in front of the sternum, held inside a life jacket and supported by a structure in the rear behind the spine. Ballistic gel was used to dampen the impact as an attempt to mimic the flesh inside the chest cavity.

Data acquisition includes the pressure and flow rate. Measurement of the pressure was performed by using fiber optic pressure sensors (Fiso FOP-M260, Canada, working range -300 mmHg to 300 mmHg, accuracy ± 3 mmHg). Placement of the pressure probe was set slightly above the aortic root, introduced through a catheter located at the same location as the brachiocephalic artery. Sampling rate of the pressure was set to 3750 Hz with a low-pass filter set at 15 kHz as recommended by the user manual for low time events (e.g., impact). Prior to any testing, calibration of the pressure waveform was done by obtaining systolic and diastolic pressures as close to 120 mmHg and 80 mmHg respectively.

Flow rate measurements were taken using two different flow meters. A magnetic in-line flow meter (ProSense® FMM75-1002, Germany, accuracy ± 0.26 L) was kept downstream from the aorta and measured the average flow rate; this flow meter was kept in the system during impact. A second flow meter (Transonic Systems INC T206) was used to measure the flow rate waveform at the aortic root at a sampling rate of 1000 Hz, scale factor of 20 L/min/V. This flow meter uses ultrasound and was only used to measure flow rate with no impact as it was sensitive to movement and would have been at the location of impact.

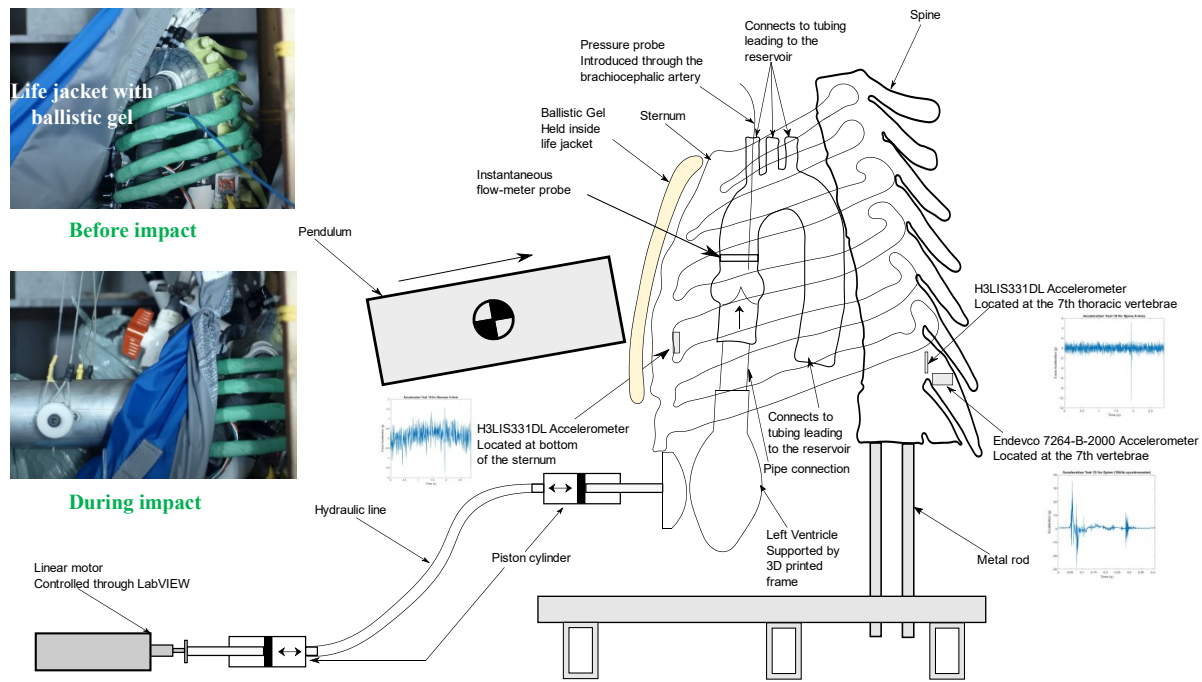

**Figure 1.** Schematic of the experimental setup. Top left actual photos of the experimental setup before and during impact.

**Table 1.** Summary of experimental parameters.

| Working Fluid | | Simulator | |
|---|---|---|---|
| Water:glycerol volume ratio | 60:40 | Heart Rate (bpm) | 70 |
| Density ($\frac{kg}{m^3}$) | 1100 | Cardiac Output (L/min) | ~5 |
| Dynamic viscosity (Pa·s) | 0.0042 | Baseline Systolic Pressure (mmHg) | ~120 |
| *In vitro materials* | | Baseline Diastolic Pressure (mmHg) | ~80 |
| Aorta | Silicone | Aortic valve diameter (mm) | 25 |
| Left Ventricle | Silicone | Mitral valve diameter (mm) | 25 |
| Spine | 3D printed PLA | **Measurements** | |
| Sternum | 3D printed PLA | 1x Pressure probe at aortic root | |
| Ribs | 3D printed PLA | 1x Flow probe at ascending aorta | |
| Flesh | Ballistic Gel | 1x Magnetic flow meter downstream of aorta | |
| **Tested Energy Input (J)** | | | |
| 7.604 | 21.062 | 35.153 | 68.405 |

Crash tests were performed using multiple masses of pendulums with adjustable heights as well as a 4K resolution 1000 Hz high-speed camera to record the impact. Testing was done with the pendulum of mass 3.8 kg. Three calibration tests were done with tap water at a height of 1229 mm from the ground to ensure that the system is leak free to prevent any potential damage to the equipment. After operation of the system was deemed satisfactory, the tap water was drained completely, and the water-glycerol mixture was used. Testing was redone at the same height of 1229 mm three times. Tests were repeated three times for heights of 1590 mm, 1968 mm and 2860 mm from the ground. Table (1) provides a summary of all the experimental conditions.

## RESULTS

Figure 2 shows examples of pressure waveforms obtained during the experiments for two different energy inputs of 7.604 J and the maximal energy of 68.405 J. The low energy impact occurred during systole. It clearly appears that it has a modest impact on the aortic pressure with a peak pressure reaching 136 mmHg. The high energy impact leads, in contrast, to a dramatic rise in the pressure up to 287 mmHg despite occurring during diastole. Furthermore, our results show that it requires more than one cycle for the pressure to recover to its normal baseline values following the impact.

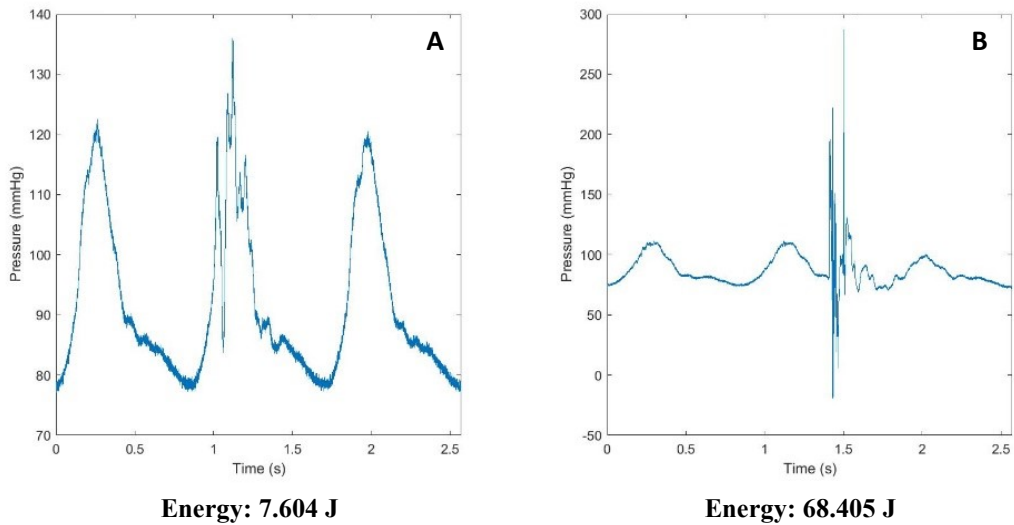

**Energy: 7.604 J**         **Energy: 68.405 J**

**Figure 2.** Instantaneous aortic pressure variations due to an impact of (A) 7.604 J and (B) 68.405 J. Note the difference in y-axis scale.

Figure 3 summarizes the results for the effect of the energy input from the pendulum on the main characteristics of the aortic pressure waveform (peak pressure, mean pressure and minimal pressure). One can note that both peak and minimal pressures are significantly affected by the change in the energy input. Interestingly, the mean pressure does not appear to be significantly affected by the energy input. This shows that following the perturbation, the overall flow recovers quite rapidly.

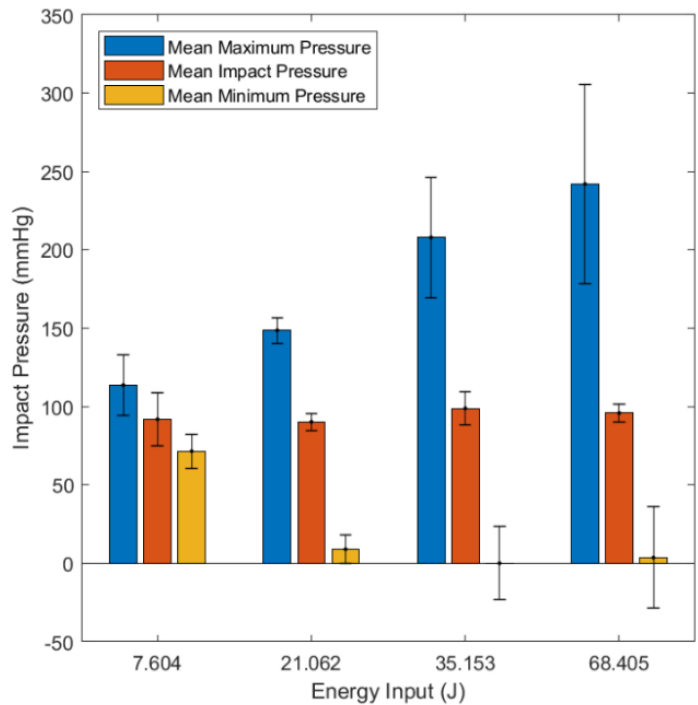

**Figure 3.** Variations in peak systolic, mean and diastolic pressures as a function of the impact energy. The reported values represent the average over the three recordings for each case.

**Figure 4** displays an interesting result regarding the effect of the timing of the impact (diastole *vs.* systole) on aortic pressure waveforms for the same impact energy. One can notice that an impact occurring during systole appears to more significantly affect the maximal peak pressure compared to an impact occurring during diastole.

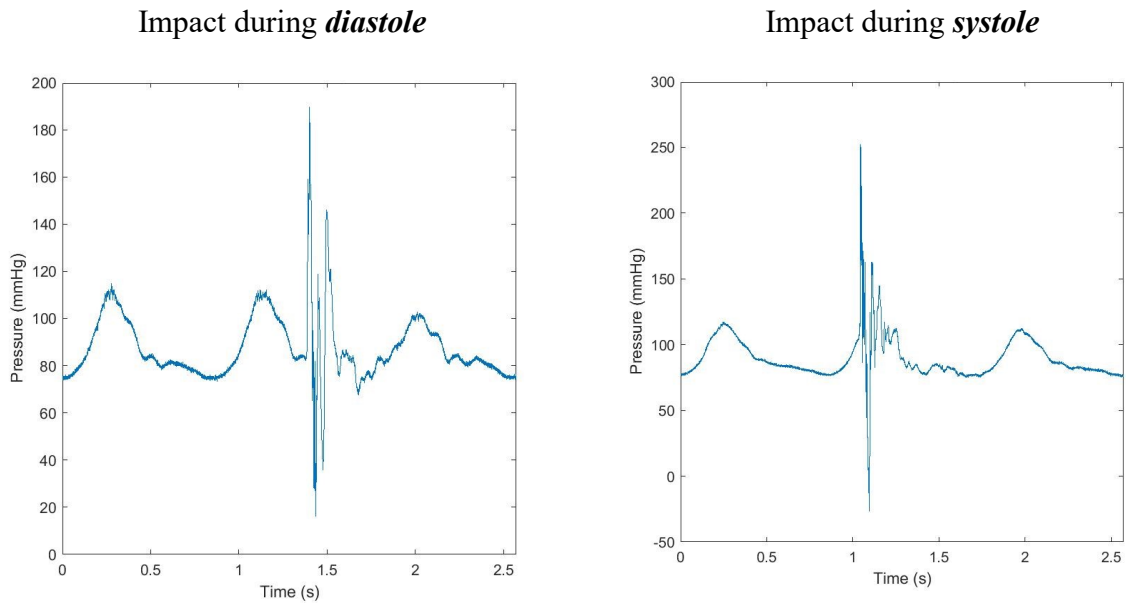

**Figure 4.** Effect of impact timing (diastole (left panel) *vs.* systole (right panel)) on aortic pressure waveforms. Note the difference in y-axis scale.

## DISCUSSION AND CONCLUSION

This work exploratory represents the first study aiming at evaluating, *in vitro* with realistic physiological pulsatile flow conditions, the effect of a sudden thoracic impact on aortic pressure waveforms. The main results show that: (1) the highest impact energy levels obtained in this study (68.405 J) can lead to peak instantaneous aortic pressures reaching values close to 2.5 times a normal systolic pressure and (2) the timing of the impact (systole *vs.* diastole) appears to significantly affects peak impact aortic pressures.

This work showcases a unique experimental model allowing to fundamentally investigate the aortic pressure response to a sudden thoracic impact and risks of blunt traumatic aortic injury. While the experimental setup introduced in this work is quite simplified, it has the merit of providing essential information that are hardly obtainable using other approaches. Studies performed with cadavers provide a high anatomic fidelity but lack the dynamic aspect of a pumping heart.

The results show that a sudden impact generates a sudden rise and drop in aortic pressure. Some values of the minimal pressure under high impact energy values can even end-up below zero (gage pressure) as probably a result of the induced water-hammer effect. Although this has still to be thoroughly confirmed by future experiments specifically designed to address this issue. The waveform then takes a moment to recover before returning to its steady-state shape. Longer time is required for waveform recovery for more severe impacts. As one would expect, higher pressures are obtained for higher energy inputs. Indeed, the maximum pressure obtained under the largest energy input of the pendulum was measured to be approximately 300 mmHg for the highest energy input of 68.405 J. This amount of pressure is significantly larger than the steady-state systolic pressure, but it is still far from the values reported to cause an aortic rupture. Klotz and Simpson, as cited by Richens et al. [1], have reached a pressure of $1.4 \times 10^5$ Pa, which corresponds to approximately 1050 mmHg. However, it is important to note that Klotz and Simpson tests were performed using cadaveric aortas that were simply pressurized at different pressures. The effect of the dynamic physiological change in the aortic pressure was not evaluated and requires more attention in the future.

Our preliminary results also suggest significant differences in peak pressures depending on the timing of the impact with higher pressures obtained during systole while the aorta is already subjected to a higher flow pressure compared to impacts during diastole. The specific contribution of the timing of the impact on the risks of blunt traumatic aortic injury requires more attention and future specific experimental studies specifically designed to address this important point.

Finally, the system and results presented in this work could help contribute towards a better understanding of the complex mechanisms leading to blunt traumatic aortic rupture and the development of preventive measures and guidelines.

**CONFLICTS OF INTEREST**

None to declare.